\documentclass[journal]{IEEEtran}
\IEEEoverridecommandlockouts
\usepackage{cite}
\usepackage[english]{babel}
\usepackage[T1]{fontenc}
\usepackage[utf8]{inputenc}
\usepackage{booktabs}
\usepackage{multirow}
\usepackage{multicol}
\usepackage{graphicx}
\usepackage{xcolor}
\usepackage{amssymb}
\usepackage{subcaption}
\usepackage{array}
\usepackage{amsmath,amsfonts,physics}
\usepackage{lipsum}
\usepackage{float}
\usepackage{balance}
\usepackage[figurename=Fig.]{caption}
\usepackage{url}
\ifCLASSINFOpdf
\else
\fi 
\begin{document} 
\title{A Neural Network--Based Energy Management System for PV--Battery Based Microgrids}
\author{Yusuf~Gupta,~\IEEEmembership{Member,~IEEE,} Mohammad~Amin,~\IEEEmembership{Senior Member,~IEEE} % <-this  stops a space
\thanks{The authors are with the Department of Electric Power Engineering, Norwegian University of Science and Technology, 7491 Trondheim, Norway (e-mail: yusuf.s.gupta@ntnu.no, mohammad.amin@ntnu.no).}% <-this % stops a space
}
\markboth{}%
{Shell \MakeLowercase{\textit{Gupta et al.}}: A Neural Network-Based Energy Management System for PV--Battery Based Microgrids}

\maketitle
\begin{abstract}
~A neural network-based energy management system (NN-EMS) has been proposed in this paper for islanded ac microgrids fed by multiple PV--battery based distributed generators (DG). The stochastic and unequal irradiation results in unequal PV output, which causes an unequal state-of-charge (SoC) among the batteries of the DGs. This effect may cause the difference in the SoCs to increase considerably over time, leading to some batteries reaching their SoC limits. These batteries would no longer be able to control the dc-link of the hybrid grid forming DG. The proposed NN-EMS ensures SoC balancing by learning an optimal state-action mapping using the outputs of an optimal power flow (OPF). The training dataset has been generated by executing a mixed-integer linear programming based OPF for droop-based island microgrids considering a practical generation-load profile. The resultant NN-EMS controller inherits the information of optimal states and the network behaviour. Compared to traditional time-ahead centralized methods, the proposed strategy does not require accurate generation-load forecasting. Further, it can also respond to the variations in the PV power in near-real-time without resorting to solving an OPF. The proposed NN-EMS controller has been validated by case studies on a CIGRE LV microgrid containing PV-battery hybrid DGs. The proposed concept can also be extended to synthesize decentralized controllers that can cooperate among themselves to achieve a global objective without communication.
\end{abstract}
\section{Introduction}
%\IEEEPARstart{M}{icrogrid}
A microgrid is an electrical network containing distributed energy resources such as distributed generators, storage and loads. It can operate as a subset of a larger electrical network or as an energy island. Grid forming DGs are beneficial for operating island microgrids with significant penetration of power-electronic interfaced energy sources. Hybrid DGs consisting of photovoltaic (PV) arrays and battery packs can be controlled as grid forming units, and they can also provide a range of power-energy services when the microgrid operates as an island \cite{saadi}. 

Hierarchical control is a popular architecture for the control of microgrids. In this structure, the control objectives are segregated into different control layers and implemented at different time scales. For instance, an optimal energy management system (EMS) is executed in the tertiary control layer every few minutes \cite{jose}. In contrast, the instantaneous power control and proportional power sharing among the DGs are implemented in a decentralized manner inside each DG. 

The focus of this paper is on the optimal power flow (OPF) executed in an EMS for island microgrids. Researchers have investigated OPFs for microgrids extensively using centralized, distributed and decentralized EMS \cite{optireview}. The OPF can be programmed using linear or nonlinear formulations and solved using conventional solvers, heuristics, or advanced learning assisted solvers \cite{ruan}. 

The OPF can be formulated to improve a technical (e.g. voltage regulation), economical (e.g. operating cost) or environmental objective. For microgrids with hybrid PV-battery based DGs, balancing the state-of-charge (SoC) of the batteries of all the DGs is vital for increased availability and reliability. In the absence of appropriate SoC balancing schemes, some batteries may reach their minimum or maximum SoC sooner than others. Furthermore, existing control strategies which aim for proportional active power sharing may not be appropriate for continuously operating the hybrid DGs in grid forming mode. Thus, it is vital to investigate advanced control strategies for balancing SoC among the batteries and maintaining acceptable power sharing.
\begin{figure*}[h]
  \centering
  \includegraphics[width=0.85\linewidth]{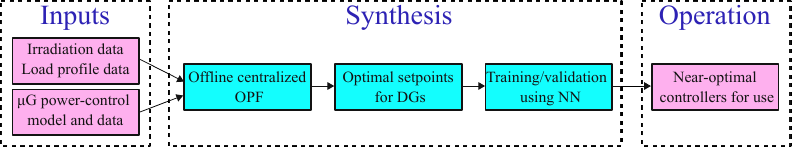}
  \caption{Overview of the steps for obtaining NN-EMS}
  \label{conceptblock}
\end{figure*}   

The control strategies and the EMSs available in the literature for island microgrids can be broadly classified into decentralized \cite{hisham1,karimi,thales,chen,lu2,hisham2,nestor,hisham4,hisham5,thales2}, distributed \cite{chendan,demin,zheng}, centralized \cite{wu,guan,gupta1,gibran,cingoz,gao}. The popular decentralized techniques are based on adaptive droop, online perturbations in power, adaptive virtual impedance, and intelligent control \cite{thales2}. Most of these strategies apply to standalone energy storage or do not consider the renewable energy and storage portion at all. Moreover, their performance is far from optimal as compared to centralized controllers. In contrast, the centralized or distributed control strategies can optimally coordinate hybrid DGs, but they operate at a slower time scale. However, the changes in PV irradiation occur at a much faster rate, often on a seconds time scale, resulting in unequal SoC on a longer time horizon. Moreover, these strategies are model-dependent and also depend on the accurate generation-load forecast. Furthermore, they also require extensive measurements from all over the network. Thus, there is a need to investigate advanced SoC balancing strategies to address these limitations. These strategies should respond faster, do not rely on up-to-date models or forecast information during operation and do not impose heavy computational burden or measurements. Artificial intelligence (AI) techniques based on machine learning (ML) and neural networks (NN) are well suited for deriving these controllers.

AI-based tools are being explored for power systems applications \cite{mahdi}. The reference \cite{trivedi} provides an overview of AI applications for microgrids in particular. There have been some efforts in using AI tools to improve the OPF formulation and reach optimal solutions quickly \cite{ruan}. These OPFs or the conventional OPFs can be employed for obtaining optimal operation datasets. This dataset can then be used for obtaining a controller which can mimic the performance of the OPF \cite{gao,kolluri,chanaka,dobbe,chen2,dong,karag}. Inspired by this philosophy, a NN based EMS (NN-EMS) has been proposed in this paper focusing on the SoC balancing aspect.

The broader steps of the proposed approach have been depicted in Fig. \ref{conceptblock}. A multi-input, multi-output (MIMO) central controller has been synthesized by training a NN with a dataset obtained by executing a mixed-integer linear programming (MILP) based OPF. The key contributions of this paper are as follows:
\begin{enumerate}
\item A NN-EMS has been proposed where NNs are employed to learn the nonlinear mapping between the cause (irradiation changes, load changes) and the control action [nominal frequency setpoints in the $P-f$ droop control] for balancing the SoC in near-real-time. The trained NN based central controller does not require heavy computations for predicting near-optimal control actions in near-real-time.
\item A MILP based OPF has been formulated for a droop controlled microgrid containing hybrid PV-battery DGs. This OPF is executed considering real generation-load profile and a dataset mapping the cause-action has been obtained for training the NN-EMS.
\item The OPF is executed such that the droop effect between the active power of the DG and the network frequency is retained. This step is crucial for generating a functional dataset for training.
\item Finally, for increased availability and reliability, decentralized intelligent controllers are also introduced, which can be useful if the central controller becomes unavailable.
\end{enumerate}

The rest of the paper is organized as follows. Section \ref{power} describes the system configuration. Section \ref{concept} presents the concept of the proposed NN-EMS structure, the OPF formulation and the steps required in synthesizing the centralized NN-EMS. Section \ref{results} contains the results and discussion, followed by the conclusions.
\section{System Configuration}
\label{power}
\begin{figure}[H]
  \centering
  \includegraphics[width=\linewidth]{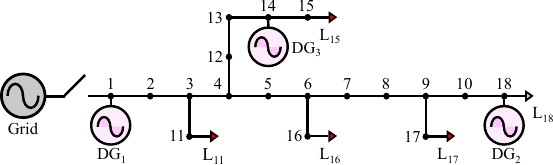}
  \caption{A typical island ac microgrid: CIGRE LV microgrid}
  \label{cigre}
\end{figure}

\begin{figure}[H]
  \centering
  \includegraphics[width=\linewidth]{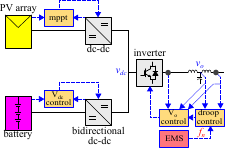}
  \caption{Power and control block diagram of a hybrid DG}
  \label{hybriddg}
\end{figure}

An island ac microgrid consisting of multiple DGs and loads, as shown in Fig. \ref{cigre}, has been considered in this paper. Each DG consists of an inverter fed by a parallel combination of a PV array and a battery pack, as depicted in Fig. \ref{hybriddg}. The PV array consists of series-parallel strings of PV panels connected to the dc-link through a dc-dc converter. Similarly, the battery pack consists of a series-parallel combination of battery modules connected to the dc-link through a dc-dc converter. 

The battery converter controls the dc-link voltage and the PV operates in maximum power point tracking (MPPT). The inverter operates in grid forming mode using the traditional $P-f$ and $Q-V$ droop controllers shown in (\ref{pdroop}) and (\ref{qdroop}), respectively.  
\begin{equation}
f_i^{*} = f_{n-i} - m_{p-i}P_{inv-i}
\label{pdroop}
\end{equation}
\begin{equation}
v_{o-i}^{*} = V_{n-i} - n_{q-i}Q_{inv-i}
\label{qdroop}
\end{equation}
where $f_i^{*}$ and $v_{o-i}^{*}$ are the reference frequency and voltage magnitude, respectively for the i$^{th}$ DG's inverter. $f_{n-i}$ and $V_{n-i}$ are the nominal setpoints of the frequency and the voltage magnitude, $m_{p-i}$ and $n_{q-i}$ are the droop slopes, and $P_{inv-i}$ and $Q_{inv-i}$ are the active and reactive power output of the inverter, respectively. The reference voltage vector generated by the droop controller is tracked by the high-bandwidth inner voltage-current control loops of the inverter.   
 
In a hierarchical control structure, $V_{n}$ and $f_{n}$ are updated by a centralized or a distributed controller at a slower time scale. The focus of the proposed strategy is to infuse the knowledge of the network, generation-load variations and the optimal operating conditions in a controller which can mimic the performance obtained from a centralized OPF in near-real-time. Although the proposed approach can be applied to a variety of objectives such as power sharing and voltage control, the focus will be on the objective of balancing the SoC of the batteries. 

The physical phenomenon can be described as follows. In the absence of a suitable SoC balancing strategy for the DGs and with fixed droop parameters, unequal PV powers would cause a proportional charging/discharging of their batteries, resulting in unequal SoC. However, by adaptively varying the nominal frequency setpoint ($f_n$) of all the DGs, a relative difference in PV powers can be transferred on the inverter side rather than changing the battery power outputs. In this way, the batteries' SoCs can be kept equal, and they will only respond to a long term generation-load imbalance. Furthermore, the DGs can keep operating in grid forming mode as long as the batteries have sufficient capacity to control the dc bus. It is to be noted that there exists a trade-off between SoC balancing and power sharing. 
\section{Proposed NN-EMS}
\label{concept}
This section explains the overall control philosophy and then details the OPF and the steps needed to synthesize the NN-based controller, which can predict optimal settings of $f_{n}$ for SoC balancing. The OPF formulation is used to obtain a dataset containing optimal operating points for various generation-load scenarios. This dataset is subsequently used to train the NN for synthesizing a centralized controller in an offline manner. In effect, an intelligent mapping is obtained between the inputs, states and the control variable.
\subsection{Control Philosophy}
\label{NN}
The SoC of the batteries are kept equal by transferring the differential amount of the PV power of the DGs to the inverter output side rather than charging or discharging the batteries unequally. Moreover, the difference between the available generation and the load demand also affects the SoC over time. Thus, an adaptive control of the inverter output powers is required, which responds to the varying generation and demand. 

\begin{figure}[h]
  \centering
  \includegraphics[width=\linewidth]{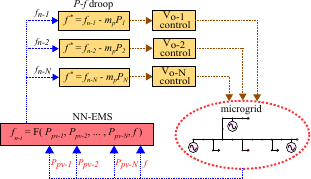}
  \caption{Working of the NN-EMS}
  \label{nncentral}
\end{figure}

It is to be noted that synthesizing such a nonlinear adaptive controller is not trivial. In this paper, a NN has been trained to obtain an intelligent centralized controller, as depicted in (\ref{pred}).  
\begin{equation}
f_{n-i}  \; =  \; F(P_{pv-1}, \; P_{pv-2}, \; ..., \; P_{pv-n}, \; f)
\label{pred}
\end{equation}
The inputs of the NN are the PV output power ($P_{pv}$) of all the DGs and the microgrid frequency ($f$). The outputs of the NN or the predicted quantities are the nominal frequency setpoints ($f_{n}$) for all the DGs. Once the NN learns the input-output relation, it can replace the traditional OPF to predict the $f_{n}$, as shown in Fig. \ref{nncentral}.
 
The choice of $P_{pv}$ and $f$ as inputs for predicting $f_{n}$ is strategic. Firstly, the SoC imbalance will happen whenever there is a difference in the irradiation ($Irr$) incident on the PV of the DGs, and the inverters still keep sharing active power as dictated by a fixed droop setting. Hence, $P_{pv}$ is the most important input for the NN. Secondly, the load also keeps varying and these variations have little or no correlation with $P_{pv}$ variations. However, the load variations are reflected in the instantaneous frequency of the microgrid due to the use of the $P-f$ droop. Hence, $f$ is taken as an additional input for training the NN. 

As depicted in Fig. \ref{nncentral}, the frequency measurement also acts as feedback to the NN-EMS while the microgrid is operating. This feedback helps the NN-EMS predict suitable $f_{n}$ setpoints and converge towards an equilibrium iteratively. Although this phenomenon works well, as will be demonstrated in the results section, the controller's optimality, convergence and stability guarantees are challenging to prove theoretically. These aspects are out of the scope of this paper and can be investigated in future.     
\subsection{OPF formulation}
This subsection describes the key equations of the OPF, which will be used to generate the training dataset for the NN-EMS. The MILP based OPF proposed in \cite{gupta1} has been used as a base, and it has been enhanced for our application. For brevity and ease of understanding, the formulation of \cite{gupta1} has not been discussed in this paper. The readers can refer to it for the OPF portion related to the network equations (KCL, KVL), load modelling, droop control, inverter output power and line loss calculation.  

The fixed inputs are the network parameters (line impedances and the rated load values), PV and battery ratings, initial SoC and the droop control parameters. The variable inputs for each iteration are the forecasted irradiation and load demand. The outputs from the OPF for further use are $P_{pv}$, $P_{bat}$, $P_{inv}$, $f$, $f_{n}$, and SoC values.
\subsubsection{\textbf{PV}}
The PV is assumed to operate in MPPT. The MPPT is assumed to operate ideally and track the MPP dictated by the $P_{pv}$ versus irradiation ($Irr$) curves. These curves are usually available from the PV manufacturer's datasheet. This allows us to employ a simple linear relation, depicted in (\ref{pvfit}), to calculate the output power from a PV panel for varying irradiation. 
\begin{equation}
P_{panel-i} \; = \; c_1 Irr_i \; + \; c_2
\label{pvfit}
\end{equation}
where $P_{panel-i}$ and $Irr_i$ are the PV panel output in W and the irradiation in W/m$^2$ for the i$^{th}$ DG. The coefficients $c_1$ and $c_2$ are the slope and the y-intercept, respectively, of the fitted linear relation. It should be noted that during night-time when $Irr$ is zero and if $c_2$ is negative, then (\ref{pvfit}) will provide a negative value for $P_{panel}$, which is impractical. To avoid this, the $P_{panel}$ value must be set to zero when $Irr$ = 0. The total $P_{pv}$ value is then determined by (\ref{Ppv}). 
\begin{equation}
P_{pv-i} \; = \; P_{panel-i} N_{pv-i}
\label{Ppv}
\end{equation}
where $N_{pv}$ is the number of panels in the PV array.  
\subsubsection{\textbf{Battery}}
The dc-dc converter and inverter are assumed to be lossless. Thus, the instantaneous power balance inside the DG can be expressed as follows:
\begin{equation}
P_{inv-i} \; = \; P_{bat-i}  + \; P_{pv-i}
\label{pbal}
\end{equation}
where $P_{bat-i}$ is the battery power output for the i$^{th}$ DG. A positive value of $P_{bat-i}$ indicates that the battery is discharging and vice versa. The SoC of the battery pack varies with time depending on the generation-load balance. The SoC percentage can be determined as follows: 
\begin{equation}
SoC_{i-k} = \frac{C_{bat-i-(k-1)} - \frac{P_{bat-i}}{V_{bat-i}} \Delta T}{C_{bat-i-rating}} \times 100
\end{equation}
where $C_{bat-i-(k-1)}$ is the available charge in the battery at the end of $(k-1)^{th}$ time step and $C_{bat-i-rating}$ is the rated charge capacity. 

The optimization objective is to balance the SoC of all the DGs for each time step. In other words, the objective is to minimize the absolute value of the sum of $dSoC$, where $dSoC$ is given by (\ref{dsoc}).   
\begin{equation}
\label{dsoc}
dSoC_{i-j-k} = SoC_{i-k} - SoC_{j-k} 
\end{equation}
where $SoC_{i-k}$ and $SoC_{j-k}$ are the state of charge of the \textit{i}$^{th}$ and \textit{j}$^{th}$ DG, respectively, for the \textit{k}$^{th}$ time step. Thus, the objective is to minimize $N_{DG}\times N_{DG}$ terms of $dSoC_{i-j}$ for each time step. The objective function can be written as follows:
% check if dSoC lower limit should be negative. very important.
\begin{equation}
\label{optieq}
\begin{aligned}
minimize \;  \sum dSoC_{i-j} \\
\end{aligned}
\end{equation}
\subsubsection{\textbf{Inverter}}
The SoC of all the DGs can be maintained equal if $P_{bat}$ remains equal for all the DGs. It is evident from (\ref{pbal}) that for unequal values of $P_{pv}$ for different DGs, their $P_{bat}$ can be maintained equally by adaptively controlling the $P_{inv}$. For example, for an increase in $P_{pv}$ for a DG, its $P_{inv}$ can be increased by a suitable amount compared to other DGs whose $P_{pv}$ are lower. Similarly, the $P_{inv}$ can be decreased in response to a decrease in $P_{pv}$. However, the change in $P_{pv}$ is not the same for all the DGs. Moreover, the change in $P_{pv}$ is stochastic. An offline OPF can consider such unequal and stochastic irradiation changes and provide optimal setpoints required for $P_{inv}$ for all the DGs. 

It should be noted that the inverters are assumed to operate in grid forming mode. Hence, the required change in $P_{inv}$ can be effected by changing the $f_{n}$ value appropriately. Thus, the SoC balancing can be achieved by dynamically predicting and controlling the $f_{n}$ values in response to a change in irradiation and load. However, the active power sharing would get disturbed if the SoC needs to be balanced. Thus, a trade-off exists between the SoC balancing and the active power sharing. An alternative is to constrain the error in the active power sharing to a tolerable limit, $\epsilon$, inside the OPF. For example, for equally rated DGs, the following constraint can be included:
\begin{equation}
P_{i-err-max} = \vert \frac{P_{inv-i} - P_{avg}}{P_{avg}} \vert \times 100 \leq \epsilon
\end{equation}
where $P_{avg}$ is determined as follows.
\begin{equation}
P_{avg} = \frac{P_{inv-1} + P_{inv-2} +...+ P_{inv-N_{DG}}}{N_{DG}}
\end{equation}
Furthermore, the line losses, $P_{loss}$, may also worsen due to the SoC equalization process. Hence, it must also be limited; for instance to 10\%, as follows:
\begin{equation}
P_{loss} \leq 0.1 \; \sum \; P_{load-rating}
\end{equation}
where $\sum \; P_{load-rating}$ is the sum of the rated loads of all the buses.

As will be demonstrated in Section \ref{results}, the $f_n$ limits considered while executing the OPF play a vital role in obtaining an easy-to-fit dataset for the NN. Instead of limiting the $f_n$ with a fixed lower and upper bound, these limits are set depending on the total load forecast during each OPF time step. This also helps preserve the droop effect between the active power and the frequency. 
\subsection{Controller Synthesis using NN}
The intelligent central controller is synthesized by training the NN with the optimal operation dataset obtained by executing the OPF for historical or forecasted irradiation and load profile. 

\begin{figure}[h]
  \centering
  \includegraphics[width=\linewidth]{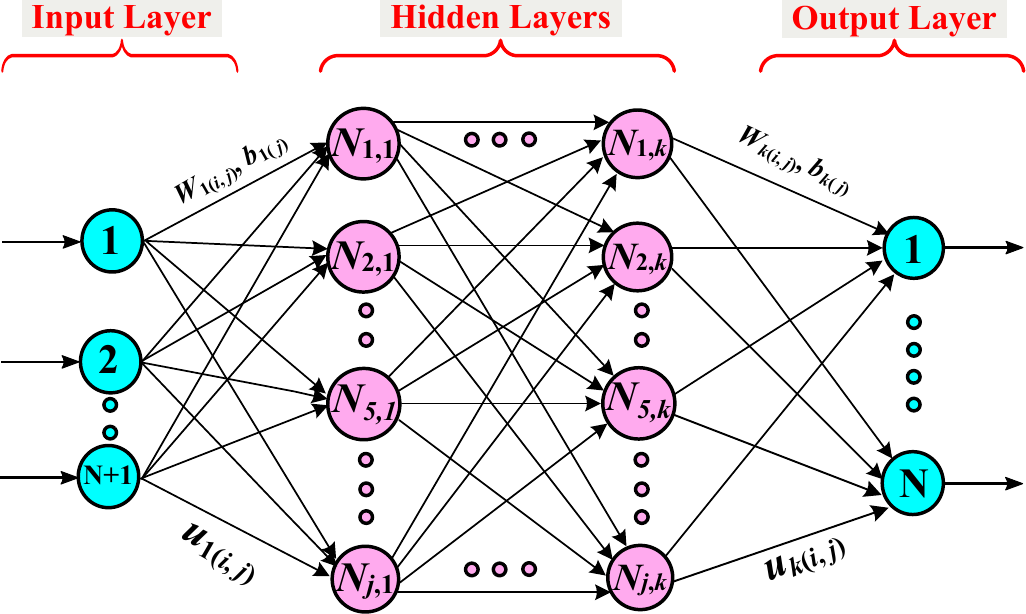}
  \caption{Structure of the employed feed-forward NN for training}
  \label{NNfig}
\end{figure}

The dataset is randomly divided into training, validation and testing sets. A two-layer feedforward NN has been used to map a relation between $N_{DG}+1$ input variables and $N_{DG}$ outputs, as shown in Fig. \ref{NNfig}. Enough number of neurons should be chosen for the hidden layer for the desired accuracy in prediction and also to avoid over-fitting. The NN adjusts the weights and biases of the hidden layer and output layer ($W_h$, $B_h$, $W_o$, $B_o$) during training for obtaining better accuracy and generalization. The fitting accuracy between the actual and the predicted outputs can be evaluated by determining the mean square error (MSE) and the regression coefficient (R) value.  
\section{Results and Discussion}
\label{results}
% @63.418, 10.402 Trondheim coordinates
A modified CIGRE low voltage microgrid, as shown in Fig. \ref{cigre}, has been used as the test system \cite{cigre}. The rated voltage and frequency of the microgrid are 400 V ph-ph and 50 Hz. It consists of 18 buses, 17 lines, five loads ($L_{11}$ -- $L_{15}$) and three DGs (DG$_{1}$ -- DG$_{3}$). The DGs are placed on buses 1, 14 and 18 for reduced power demand, according to \cite{gupta1}. They are considered equal in rating. The rated load values have been scaled down to one-third compared to \cite{cigre}. The load values are included in Table \ref{load}.
\begin{table}[h]
\centering
\caption{Load data}
\label{load}
\scalebox{0.98}{%
\begin{tabular}{lrrrrrrrrr}
\toprule
Bus no.      & 1  & 2     & 3  & 4  & 5  & 6      & 7     & 8     & 9     \\
P (kW)       & 0  & 0     & 0  & 0  & 0  & 0      & 0     & 0     & 0     \\
Q (kvar)     & 0  & 0     & 0  & 0  & 0  & 0      & 0     & 0     & 0     \\
\toprule
Bus no.      & 10 & 11    & 12 & 13 & 14 & 15     & 16    & 17    & 18    \\
P (kW)       & 0  & 1.62  & 0  & 0  & 0  & 16.15  & 6.52  & 1.62  & 7.08  \\
Q (kvar)     & 0  & 1.00  & 0  & 0  & 0  & 10.01  & 4.04  & 1.00  & 4.39  \\
\bottomrule
\end{tabular}%
}
\end{table}

Each DG is fed by a PV array and a battery pack which have been sized appropriately considering the PV profile \cite{soldata} and load pattern \cite{loaddata} of May 2020 for the city of Trondheim in Norway. The available data have been converted from hourly values to 5-minute values by linear interpolation. To introduce different variations in the irradiation of the DGs, random gains between 1 and 1.14 have been multiplied to the PV pattern of \cite{soldata}. The resultant irradiation, $Irr_1$, $Irr_2$ and $Irr_3$ are shown in Fig. \ref{ILS}(a). The load profile is also demonstrated in Fig. \ref{ILS}(c), where the peak and the minimum load are approximately 1.5 and 0.85 times the rated total load of approximately 33 kW.

\begin{figure}[h]
  \centering
  \includegraphics[width=\linewidth]{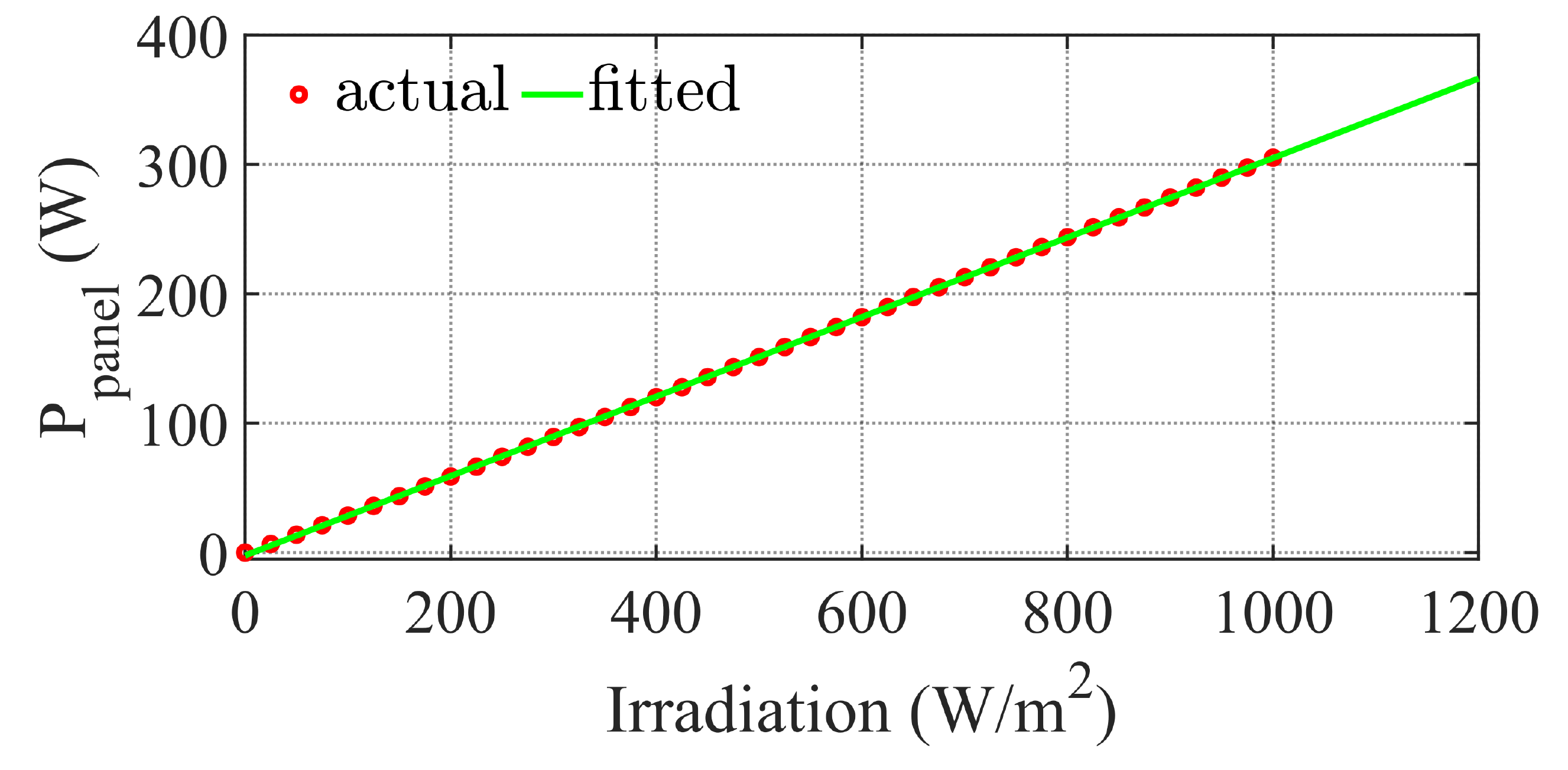}
  \caption{Actual vs fitted values for $P_{panel}$}
  \label{IP}
\end{figure}

The PV panels and the dc-dc converter have been sized to provide 70 kWp power and 650 V at the dc-link of the inverter. The PV panels from SunPower (model no: SPR-305E-WHT-D) have been considered. The power output data of the PV panel versus the input irradiation is included in Table \ref{IPtab}. The MATLAB function \textit{polyfit} has been used to fit the data from this table. The fitted line is shown in Fig. \ref{IP}.
\begin{table}[h]
\centering
\caption{$Irr$ vs $P_{panel}$ for a 305 W PV module}
\label{IPtab}
\scalebox{0.95}{%
\begin{tabular}{lrrrrrrr}
\toprule
$Irr$ (W/m$^2$)      & 0     & 25    & 50    & 75    & 100   & 125   & 150    \\
$P_{panel}$ (W)      & 0.0   & 6.7   & 13.9  & 21.2  & 28.6  & 36.1  & 43.6   \\
\midrule
$Irr$ (W/m$^2$)      & 175   & 200   & 225   & 250  & 275   & 300   & 325     \\ 
$P_{panel}$ (W)      & 51.2  & 58.8  & 66.4  & 74.1  & 81.8  & 89.4  & 97.1   \\
\midrule
$Irr$ (W/m$^2$)      & 350   & 375   & 400   & 425   & 450   & 475   & 500    \\
$P_{panel}$ (W)      & 104.8 & 112.5 & 120.2 & 127.8 & 135.7 & 143.4 & 151.1  \\
\midrule
$Irr$ (W/m$^2$)      & 525   & 550   & 575   & 600   & 625   & 650   & 675    \\
$P_{panel}$ (W)      & 158.9 & 166.6 & 174.3 & 182   & 189.8 & 197.5 & 205.2  \\
\midrule
$Irr$ (W/m$^2$)      & 700   & 725   & 750  & 775   & 800   & 825   & 850     \\
$P_{panel}$ (W)      & 212.9 & 220.6 & 228.3  & 236.1 & 243.8 & 251.5 & 259.1 \\
\midrule
$Irr$ (W/m$^2$)      & 875   & 900   & 925   & 950   & 975   & 1000           \\
$P_{panel}$ (W)      & 266.8 & 274.5 & 282.1 & 289.9 & 297.6 & 305.2          \\
\bottomrule
\end{tabular}%
}
\end{table}
\subsection{Data Collection}
The critical parameters of the OPF and the NN-EMS are included in Table \ref{emstab}. The OPF is fed with the 5-minute interval data of irradiation and load. The OPF has been written in MATLAB and solved using the IBM CPLEX solver.
\begin{table}[h]
\centering
\caption{OPF and NN-EMS parameters}
\label{emstab}
%\setlength{\tabcolsep}{1.5pt}
%\scalebox{1}{%
\begin{tabular}{rrr}
\toprule
\textbf{Parameter}                       & \textbf{Value}     &	\textbf{Unit}   \\
\midrule
$N_{pv}$  	                             & 8$\times$29        & --              \\
$c_1, c_2$ 			                     & 0.3072, $-$2.1944  &  --             \\
\midrule
$V_{bat}$ 			                     & 550                &  V              \\
%$AH_{bat}$ 			                 & 4500               &  AH             \\
$SoC_{min}$, $SoC_{max}$  	             & 10, 90             & \%              \\
$dSoC_{min}$, $dSoC_{max}$  	         & 0, 10              & \%              \\
$P_{i-err-max}$                          & 60                 & \%              \\
\midrule
$m_p$ 				                     & 0.005              & rad/s/kW        \\
$n_q$ 				                     & 0.1                & V/kvar          \\
$\epsilon$ 		                         & 60                 & \%              \\
\midrule
$V_{min}$, $V_{max}$  	                 & 0.95, 1            & p.u.              \\
$f_{min}$, $f_{max}$  	                 & 0.95, 1            & p.u.              \\
\bottomrule
\end{tabular}
%}
\end{table}
\begin{figure}[h!]
  \centering
  \includegraphics[width=\linewidth]{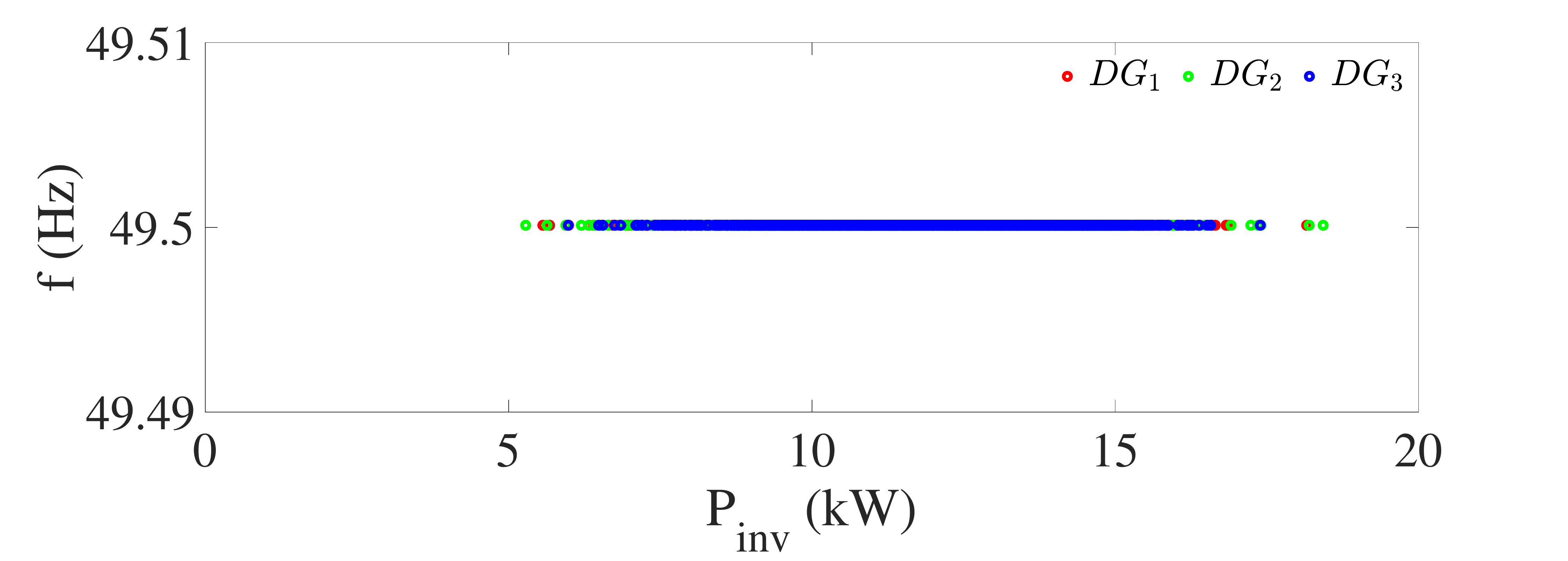}
  \caption{Droop effect lost}
  \label{nodroop}
\end{figure}
\begin{figure}[h!]
  \centering
  \includegraphics[width=\linewidth]{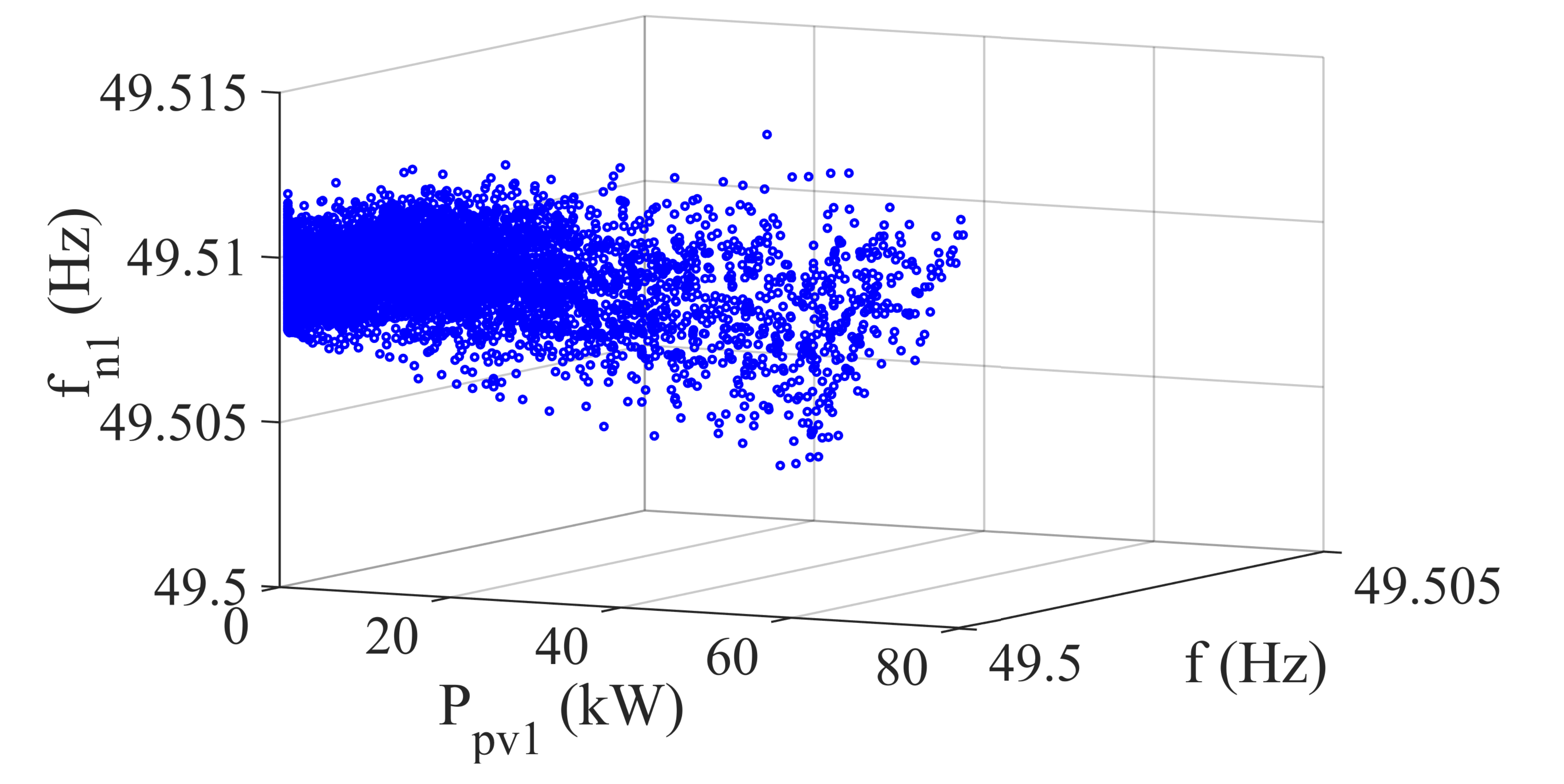}
  \caption{Plot of $f_n$ vs $P_{pv1}$ and $f$: No steps for $f_n$ limits}
  \label{3dnoflim}
\end{figure}

The OPF is first executed sequentially for the 31 days (31$\times$24$\times$12 $\rightarrow$ 8928 time steps) considering fixed bounds of 0.99--1.01 per unit (p.u.) for $f_n$. The optimal values of $f$ versus $P_{inv}$ are plotted for all the DGs in Fig. \ref{nodroop}. It can be observed that all the obtained optimal values of the frequency lie near the lower bound of 0.99 p.u. Thus, the droop effect is lost since all the $f$ values are nearly the same irrespective of the inverter output power. For better visualization, the optimal values of $f_n$ versus $P_{pv}$ and $f$ for DG$_1$ are plotted in Fig. \ref{3dnoflim}. It should be noted that it is challenging to fit a relation of $f_n$ with $P_{pv}$ and $f$ using such a dataset.  

\begin{figure}[h]
  \centering
	  \includegraphics[width=\linewidth]{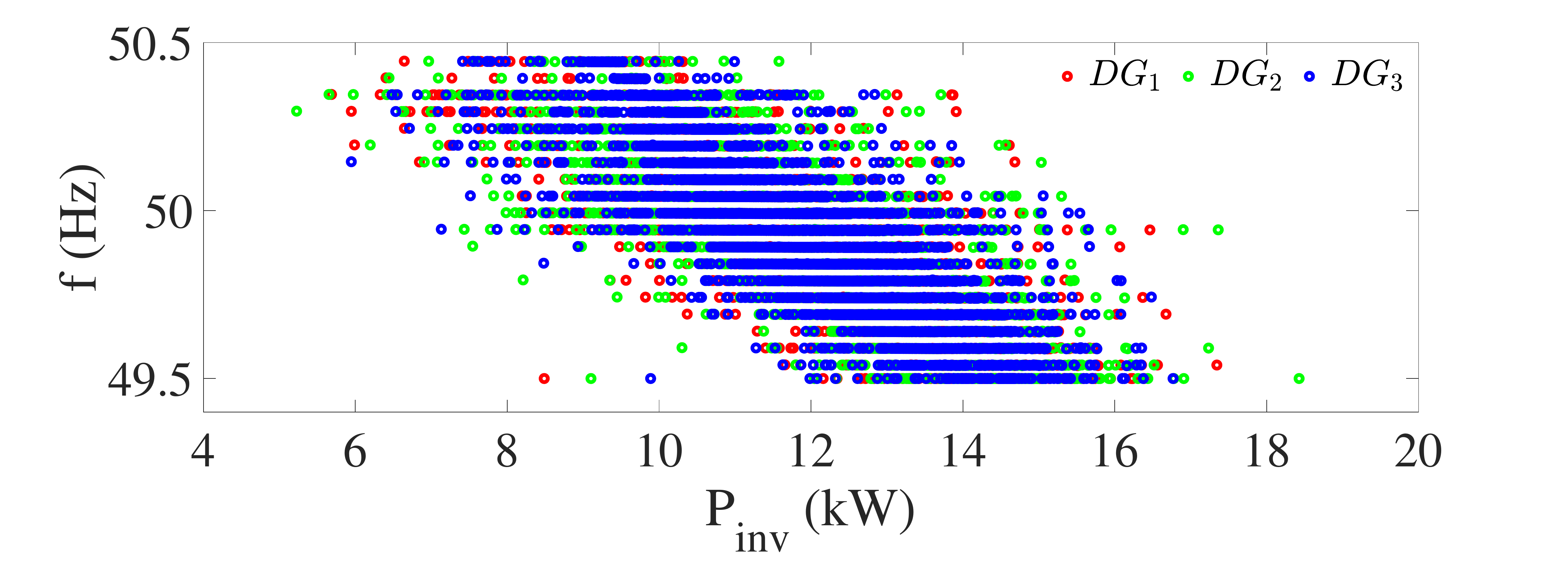}
  \caption{Droop effect preserved}
  \label{forceddroop}
\end{figure}

\begin{figure}[h]
  \centering
  \includegraphics[width=\linewidth]{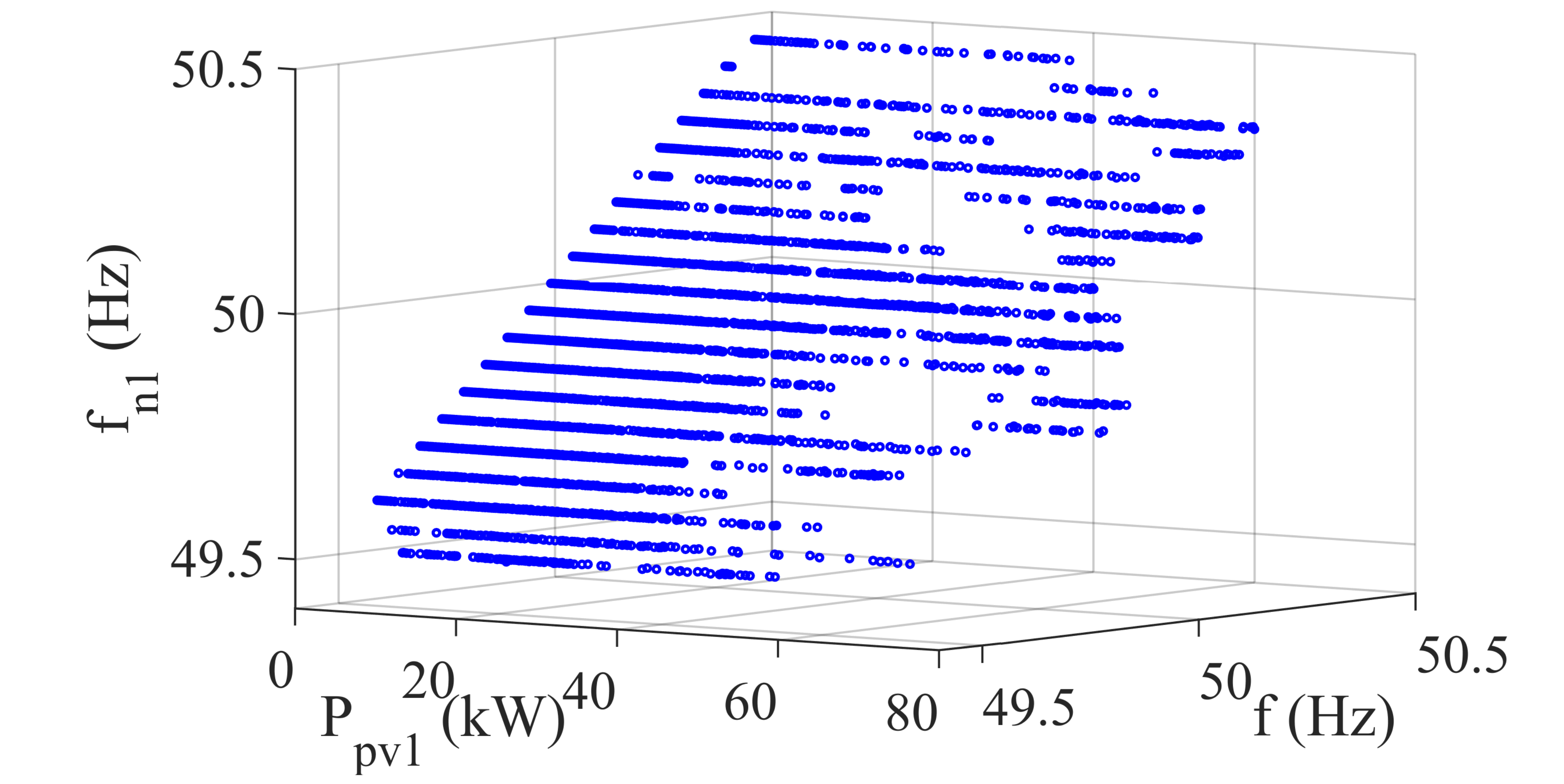}
  \caption{Plot of $f_n$ vs $P_{pv1}$ and $f$: 20 steps for $f_n$ limits}
  \label{3d}
\end{figure}

To preserve the droop effect and also to spread the $f_n$ values, the OPF was rerun with $f_n$ constrained in steps according to the p.u. value of the forecasted load. The $f_n$ upper and lower limits have been successively decreased from 1.01 to 0.99 p.u. in 20 equal steps of 0.001 p.u. For example, the $f_n$ limits are set to 1.010--1.009 p.u. if the load value is between 1 p.u. and 0.95 p.u., and it is set to 1.009--1.008 if the load value is between 0.95 p.u. and 0.90 p.u. and so on. The drooping effect is preserved, as illustrated in Fig. \ref{forceddroop}. Furthermore, as seen in Fig. \ref{3d}, the data points are spread out, thus facilitating easier and more accurate fitting by the NN in the next step.

\begin{figure*}[ht]
  \centering 
  \includegraphics[width=0.99\linewidth]{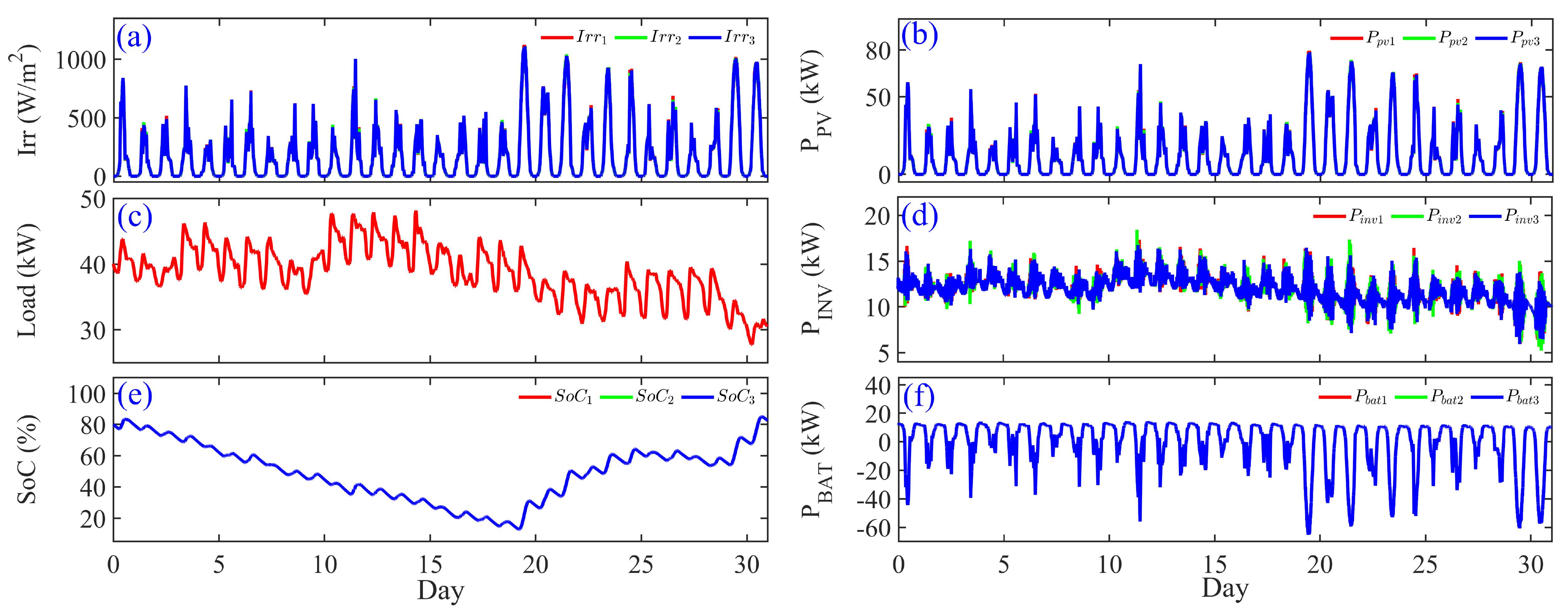}
  \caption{The inputs and outputs of the OPF for 8928 sequential runs of 5 minute time step}
  \label{ILS}
\end{figure*}

The dataset consisting of 8928 optimal steady-state operating points is saved for the training and validation steps. As seen in Fig. \ref{ILS}(b), the maximum $P_{pv}$ is about 80 kW, whereas the maximum and minimum $P_{bat}$ are 14.47 kW and --65.28 kW, respectively. The $P_{err-avg}$ and $P_{err-max}$ are found to be 2.54\% and 32.71\%, respectively, over the whole month. As seen from the SoC and $P_{bat}$ waveforms in Fig. \ref{ILS}(e) and Fig. \ref{ILS}(f), the OPF can provide a perfect SoC balancing. It should also be noted that the batteries operate in charging mode most of the time. 
%\begin{table}[h]
%\centering
%\caption{$f_n$ limits}
%\label{flim}
%%\setlength{\tabcolsep}{1.5pt}
%\scalebox{1}{%
%\begin{tabular}{lrrrrr}
%\toprule
%Load (p.u.)            & 0$-$0.1   & 0.1$-$0.2 & 0.2$-$0.3 & 0.3$-$0.4 & 0.4$-$0.5 \\
%$f_{n-max}$ (p.u.)     & 1.010     & 1.008     & 1.006     & 1.004     & 1.002 \\
%$f_{n-min}$ (p.u.)     & 1.008     & 1.006     & 1.004     & 1.002     & 1.000 \\ 
%\midrule
%Load (p.u.)            & 0.5$-$0.6 & 0.6$-$0.7 & 0.7$-$0.8 & 0.8$-$0.9 & 0.9$-$1.0 \\
%$f_{n-max}$ (p.u.)     & 1.000     & 0.998     & 0.996     & 0.994     & 0.992   \\
%$f_{n-min}$ (p.u.)     & 0.998     & 0.996     & 0.994     & 0.992     & 0.990 \\ 
%\bottomrule
%\end{tabular}%
%}
%\end{table}
\begin{figure*}[h!]
  \centering
  \includegraphics[width=0.99\linewidth]{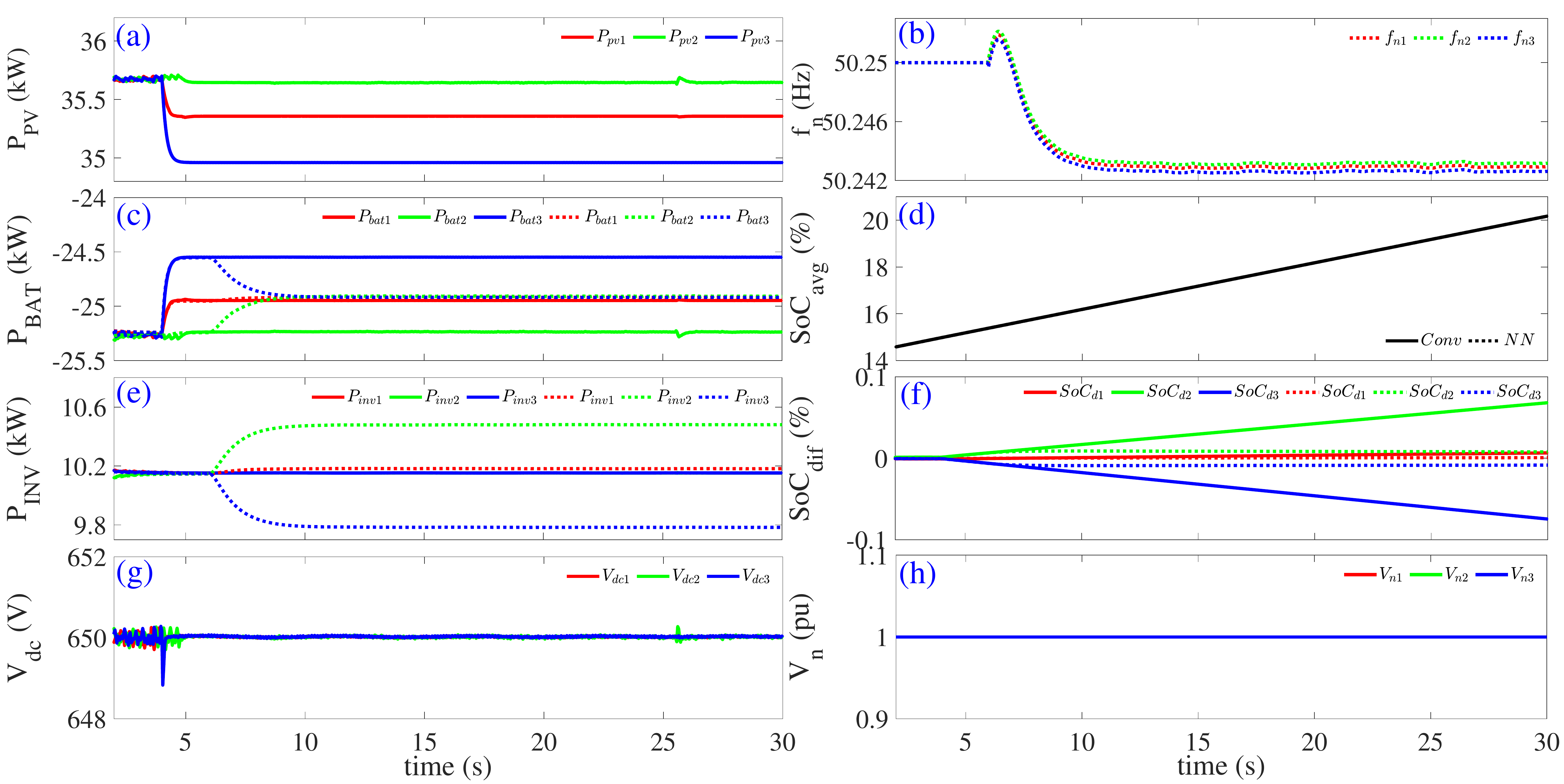}
  \caption{Comparison of the NN-EMS (dotted lines) with conventional droop (solid lines): Battery charging case}
  \label{charge}
\end{figure*}
\begin{figure*}[h!]
  \centering
  \includegraphics[width=0.99\linewidth]{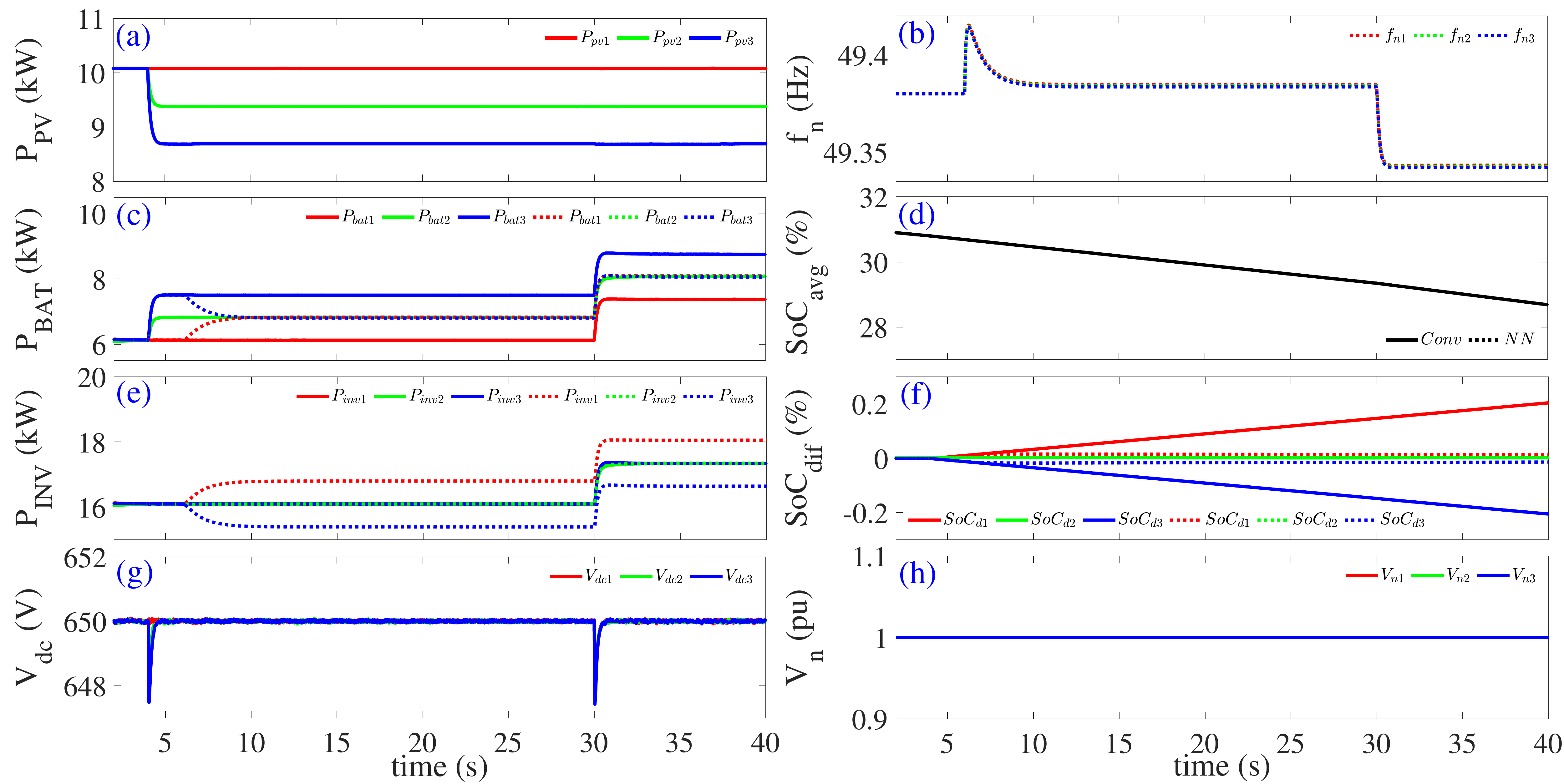}
  \caption{Comparison of the NN-EMS (dotted lines) with conventional droop (solid lines): Battery discharging case}
  \label{discharge}
\end{figure*}
%\begin{figure*}[h!]
%  \centering
%  \includegraphics[width=0.99\linewidth]{figures/load_cent.eps}
%  \caption{Comparison of the NN-EMS (dotted lines) with conventional droop (solid lines): Battery discharging case, load changed.}
%  \label{loadD}
%\end{figure*}
\subsection{Training the NN based controller}
The training has been performed using the neural network fitting application in MATLAB 2020b. The dataset has been randomly divided into 80\%, 10\% and 10\% cases for training, validation, and testing. The Levenberg-Marquardt algorithm with ten neurons has been selected for training. Since there are three DGs in the test microgrid ($N_{DG}$ = 3), the NN-EMS must be trained for mapping four inputs to three outputs. The MSE and R of the fit are close to 2e--6 and 1, respectively.  
% I can instead plot wn1-wn3: their true value and predicted value? in a 3d diagram??
\subsection{Working of the trained NN-EMS}
A Simulink model of the CIGRE microgrid has been constructed. A detailed model of the DG has been implemented. The DG consists of PV plus boost converter (MPPT), battery plus bidirectional converter (dc-link voltage control) and an inverter with LCL filter, as shown in Fig. \ref{hybriddg}. The inverter control system consists of a droop controller and an inner current-voltage-PWM controller. The simulation runs at a time step of 2.5 $\mu$s.

The NN-EMS scans the $f$ and $P_{pv-1,2,3}$ and predicts $f_{n-1,2,3}$ instantaneously. The $f_{n-1,2,3}$ setpoints are then communicated to the respective DG’s droop controller. The NN-EMS can mimic the performance of a centralized OPF on a much faster time scale. Its capability to act faster is crucial because PV variations happen on seconds scale. Furthermore, the NN-EMS needs fewer measurements compared to a centralized OPF.
\subsubsection{\textbf{Battery charging scenario}}
The NN-EMS has been tested for a generation-load combination which entails the batteries to operate in charging mode, as depicted in Fig. \ref{charge}. The NN-EMS is disabled at the start of the simulation, and the DGs run with conventional droops with a fixed nominal frequency of 50.25 Hz till 6 s. The SoC of all the batteries are assumed to be 14\% approximately. The irradiation is set to 600 W/m$^2$ at the start and the load levels are set to 1 p.u. The waveforms in Fig. \ref{charge} correspond to the microgrid's response with fixed droop and the proposed NN-EMS based adaptive droop. They have been represented by solid and dotted patterns, respectively. 

The irradiations are changed unequally at 4 s to 575 W/m$^2$, 600 W/m$^2$ and 550 W/m$^2$ for $DG_{1,2,3}$, respectively. As seen in Fig. \ref{charge}, the $P_{pv}$ waveforms are stable due to the proper working of the MPPT. The dc-link voltage ($V_{dc}$) is also seen to be tightly controlled at 650 V by the battery dc-dc converter. Thus, the power balance is working well at the dc-link. The inverters' active power sharing is equal initially due to the $P-f$ droop. It is to be noted that the $Q-V$ droop parameters have been kept constant throughout the case studies since the focus of the work is not on reactive power sharing.

It can be further observed from Fig. \ref{charge} that after 4 s, the difference in the $P_{pv}$ of the DGs causes their $P_{bat}$ output to become unequal. This triggers a divergence in the SoCs, as seen by the $SoC_{dif}$ waveforms. $SoC_{dif}$ is the deviation of the SoCs from their mean value $SoC_{avg}$ as shown in (\ref{dif}).
\begin{equation}
SoC_{dif} \; = \;SoC_i \; - \; SoC_{avg}
\label{dif}
\end{equation}

$SoC_{avg}$ is also plotted in Fig. \ref{charge}. Ideally, all the $SoC_{dif}$ values should converge to zero. However, this difference will increase over time without a suitable SoC balancing strategy. The NN-EMS is enabled at 6 s, and it starts predicting and communicating the $f_{n-1,2,3}$ setpoints to the DGs, as shown in Fig. \ref{charge}(b). The difference in $f_{n-1,2,3}$ setpoints causes the DGs to produce unequal power, as seen from the dotted waveforms of Fig. \ref{charge}(e). Thus the DGs try to extract unequal power from their dc links. This, in turn, forces their batteries to absorb unequal powers, as seen in Fig. \ref{charge}(b). It can be seen that the $P_{bat-1,2,3}$ converge towards an almost equal value. In this way, the SoCs do not diverge from each other and remain almost equal, as seen in Fig. \ref{charge}(f). This validates the effectiveness of the proposed NN-EMS in maintaining near-perfect SoC balance.     
\subsubsection{\textbf{Battery discharging scenario}}
It can be observed from Fig. \ref{ILS}(f) that the training dataset majorly contains charging scenarios. Thus, it is also essential to test the proposed technique for the battery discharging scenario, as depicted in Fig. \ref{discharge}. 

The DGs run with conventional droop with a fixed nominal frequency of 49.38 Hz till 6 s. The SoC of all the batteries are assumed to be 31\% approximately. The irradiation is set to 150 W/m$^2$ initially, and the load levels are set to 1.66 p.u. It should be noted that the NN-EMS is not trained for this load level.

The irradiations are changed at 4 s to 150 W/m$^2$, 140 W/m$^2$ and 130 W/m$^2$ for $DG_{1,2,3}$, respectively. It can be observed from Fig. \ref{discharge}(a) and \ref{discharge}(c) that due to the difference in the $P_{pv}$ of the DGs, the $P_{bat}$ becomes unequal. This triggers a sharp divergence in the SoCs, as seen in Fig. \ref{discharge}(f). 

The NN-EMS is enabled at 6 s, and it starts predicting and communicating the $f_{n-1,2,3}$ setpoints to the DGs, as shown in Fig. \ref{discharge}(b). The difference in $f_{n-1,2,3}$ setpoints causes the DGs to produce unequal power, as seen from the dotted waveforms of Fig. \ref{discharge}(e). The DGs extract unequal power from their dc links. This, in turn, forces the batteries to provide unequal powers, as seen in Fig. \ref{discharge}(b). The $P_{bat-1,2,3}$ outputs converge to almost equal value. In this way, the SoCs of the batteries remain almost equal, as seen in Fig. \ref{discharge}(f). Thus, the proposed NN-EMS works well in discharging scenarios also, although it is trained majorly for the charging scenarios.

It is crucial that the NN-EMS also provides accurate predictions even when there is a sudden change in load. To test this, a load of 5 kW and 3 kVA is added on bus 15 at 30 s. It can be observed that the $f_{n-1,2,3}$ predictions smoothly move to a new operating point in response to the load change. The SoC balance is still preserved, as seen from the near-zero values of $SoC_{dif}$. This further validates the effective operation of the NN-EMS.     
%\subsection{Performance over a day}
%Create a test day profile and compare the output of optimization, ANN prediction and conventional droop. 
\subsection{Extending the NN-EMS for decentralized implementation}
Although the centralized NN-EMS provides excellent performance, it is prone to communication delay or failure. For better reliability, the proposed control philosophy can also be extended in future to obtain decentralized NN based controllers for the DGs. For instance, each DG's controller can consider its $P_{pv}$ and the network frequency as the inputs for predicting its $f_{n}$. Thus, the following function needs to be mapped for each DG:  
\begin{equation}
f_{n-i} \; = \;  F(P_{pv-i}, \; f)
\label{preddec}
\end{equation}
\begin{figure}[H]
  \centering
  \includegraphics[width=\linewidth]{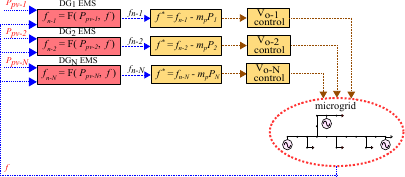}
  \caption{Decentralized implementation of a trained NN model}
  \label{nndecentral}
\end{figure}
\begin{figure*}[h!]
  \centering
  \includegraphics[width=0.99\linewidth]{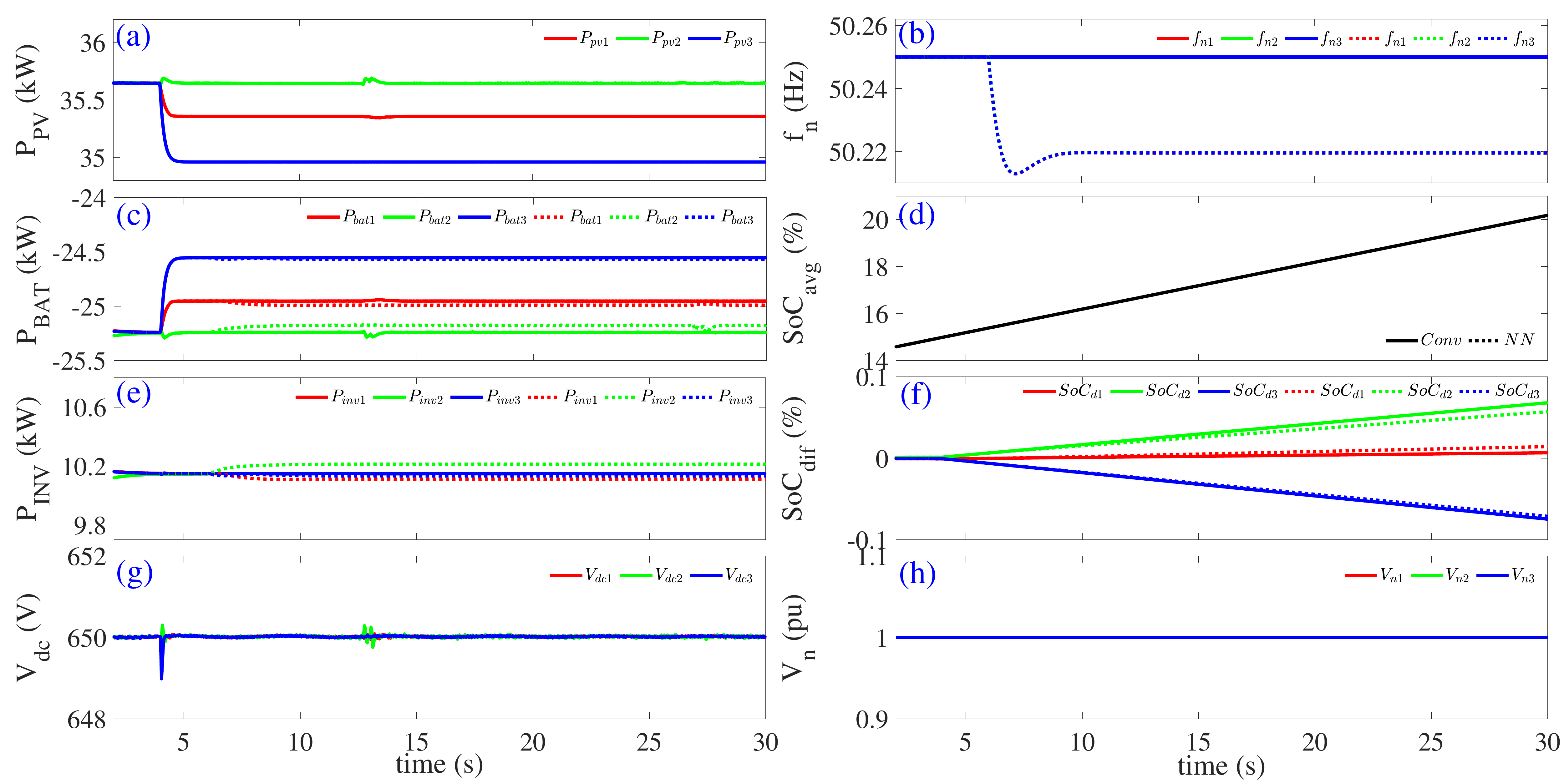}
  \caption{Comparison of decentralized NN based controllers (dotted lines) with conventional droop (solid lines): Battery charging case}
  \label{chargedec}
\end{figure*}

As shown in Fig. \ref{nndecentral}, the decentralized approach would facilitate a completely communication-less operation since the predictions happen locally in each DG using the locally available measurements. However, it is very challenging for the classical NN techniques to accurately fit (\ref{preddec}) for all the DGs and perform well during operation. 

The main problem is that the $f_{n}$ setpoints obtained from the OPF for any given generation-load scenario lie very close to each other. The proximity of the $f_{n}$ setpoints makes the NN prone to inaccurate predictions, which is undesirable. To demonstrate this, decentralized controllers were synthesized according to (\ref{preddec}), and the performance was tested for the charging scenario, as depicted in Fig. \ref{chargedec}. 

The MSE and the R obtained were close to 1e--5 and 1, respectively. Although the fitting accuracy is excellent, the performance during operation is not as expected. It can be seen from Fig. \ref{chargedec}(f) that the SoC balancing does not improve much with the decentralized controllers. In the future, advanced NN techniques like deep learning can be explored to enhance the performance of the decentralized approach.  
\section{Conclusions}
\label{conclusion} 
A neural network (NN) based centralized EMS has been proposed for microgrids containing PV-battery hybrid DGs. The NN learns the optimal control actions to be taken to maintain the SoC balance among the batteries of different DGs while the irradiation and load changes occur. The NN-EMS has been trained using optimal operating points obtained using an OPF. Thus, it inherits the information of optimal states and the network behaviour. The SoC of the batteries are maintained nearly equal by transferring the differential amount of the PV power to the inverter output side rather than charging/discharging the batteries unequally. This is achieved by adapting the $P-f$ droop controller setpoints using the trained NN-EMS. The proposed concept has been validated by case studies on a CIGRE LV microgrid and is shown to maintain almost equal SoCs. The NN-EMS predicts near-optimal setpoints in real-time without depending on an accurate generation-load forecast or solving an OPF. Further, it also responds effectively to the variations in the PV power and load changes. The proposed concept can also be thoroughly investigated in future to synthesize completely decentralized controllers that can cooperate among themselves to achieve a network-wide objective.
%\section*{Acknowledgement}
%Authors would like to acknowledge the support received from the Department of Electric Power Engineering, NTNU Norway.
\bibliographystyle{IEEEtran}
\balance
\bibliography{refEMS}
\end{document}